\documentclass[journal]{IEEEtran}
\usepackage{fancyhdr} 
\usepackage[british,UKenglish,USenglish,english,american]{babel}
\usepackage{epsfig,graphics,subfigure,graphicx,latexsym,longtable,amsmath,amscd,latexsym,amssymb,mathrsfs,syntonly,eucal}
\usepackage{cite}
\usepackage{multirow}
\usepackage[T1]{fontenc}
\usepackage{bm,cite}
\usepackage{amsbsy}
\usepackage{latexsym}
\usepackage{wasysym}
\usepackage{placeins}
\usepackage[lined,boxed,commentsnumbered]{algorithm2e}

\usepackage{bm,cite}
\usepackage{amsmath}
\usepackage{amsbsy}
\usepackage{latexsym}
\usepackage{amssymb}
\usepackage{wasysym}

\newtheorem{theorem}{Theorem}
\newtheorem{proposition}{Proposition}
\newtheorem{lemma}{Lemma}

\newtheorem{definition}{Definition}

\DeclareMathAlphabet{\mathbit}{OML}{cmr}{bx}{it}
\DeclareMathAlphabet{\mathsf}{OT1}{cmss}{m}{n}
\DeclareMathAlphabet{\mathTXf}{OT1}{cmss}{bx}{it}

\DeclareMathOperator{\Transpose}{T}

\DeclareMathOperator*{\argmax}{arg\ max}

\DeclareMathOperator{\Exp}{E}

\DeclareMathAlphabet{\mathpzc}{OT1}{pzc}{m}{it}

\newcommand{\Tr}{{\Transpose}}

\newcommand{\He}{{{\text{H}}}}

\graphicspath{{./figures/}}

\title{Optimization of Energy Harvesting MISO Communication System with Feedback}

\author{\IEEEauthorblockN{Rajeev Gangula$^{*}$, David Gesbert$^{*}$, and Deniz G{\"u}nd{\"u}z\ $^{\S}$\\}

\IEEEauthorblockA{\textsuperscript{$^{*}$}Mobile Communications Department, Eurecom, France\\
\textsuperscript{$^{\S}$}Dept. of Electrical and Electronic Engineering, Imperial College London, UK\\
Email: \text{\{gangula,gesbert\}@eurecom.fr}, \text{d.gunduz@imperial.ac.uk}
}
}

\begin{document}
\maketitle

\begin{abstract}
Optimization of a point-to-point (p2p) multiple-input single-output (MISO) communication system 
is considered when both the transmitter (TX) and the receiver (RX) have energy harvesting (EH) capabilities. 
The RX is interested in feeding back the channel state information (CSI) to the TX to help improve the 
transmission rate. The objective is to maximize the throughput by a deadline, subject to the EH constraints at 
the TX and the RX. The throughput metric considered is an upper bound on the ergodic rate of the MISO channel with 
beamforming and limited feedback. Feedback bit allocation and transmission policies that maximize the upper bound 
on the ergodic rate are obtained. Tools from majorization theory are used to simplify the formulated optimization 
problems. Optimal policies obtained for the modified problem outperform the naive scheme in which no intelligent 
management of energy is performed.
\end{abstract} 


\section{Introduction} \label{sec1}
Powering up terminals in communication networks by renewable ambient energy reduces the carbon footprint of 
the information and communication technologies, which can no longer be neglected with the exponential growth 
in the number of communication devices. Another advantage of EH technology is that, it increases the autonomy 
of battery-run communication devices. In traditional wireless networks nodes get their energy from the power 
grid by always or periodically connecting to it. While it is easy to connect the terminals to the grid in some networks, 
in others, such as sensor networks, it cannot be done once after the deployment. Therefore, in such networks 
a node's lifetime, and hence, the network lifetime, is constrained by the limited initial energy in the battery. 
Providing EH capabilities to the communication nodes is an attractive solution to the network lifetime problem \cite{Kansal2007}. 
An EH node can scavenge energy from the environment (typical sources are solar, wind, vibration, thermal, etc.) \cite{sudev}. 
With EH nodes in the network, in principle, one can guarantee perpetual lifetime without the need of replacing batteries.

However, EH poses a new design challenge as the energy sources are typically sporadic and random. The main challenge lies in 
ensuring the Quality of Service (QoS) constraints of the network given the random and time varying energy sources. 
This calls for the intelligent management of various parameters involved in a communication system.

Recently, a significant number of papers have appeared studying the optimal transmission schemes for EH communication 
systems under different assumptions regarding the node's knowledge about the underlying EH process. Offline optimization framework deals with systems in which non-causal knowledge of the EH process is available. Within this frame work, optimal transmission schemes are studied for the p2p fading channel \cite{ulkus_11}, broadcast channel \cite{elif}, \cite{ulukus_bc}, \cite{deniz_leak} and relay channel \cite{deniz_relay, haung}. 
See \cite{deniz_tuto} for an extensive overview. 

To the best of our knowledge, a common aspect of all prior works on EH communication networks is that the TX is assumed 
to have access to perfect CSI. Knowledge of the CSI at the TX is beneficial in designing the optimal channel adaptation 
techniques and the TX filters in multi-antenna systems. However, recent studies have demonstrated that, although feedback 
enhances the system performance, feedback resources, namely power and bandwidth, are limited, and must be spent wisely \cite{love_08}. 
As a result, an important question arises: How do the EH constraints affect the design of feedback enabled wireless networks?

In this paper, we study the optimization of a feedback enabled EH MISO channel, 
where feedback is used to improve the rate through array gain. The system model and the main assumptions in this paper
are given in Section III. In Section IV, we consider the optimization of the feedback policy under EH constraints at the RX, 
while the TX is assumed to have a constant power supply. The motivation is to address the following:
In the case of EH, the available energy at the RX varies over time. Should the RX feedback same quality of CSI at all times?
If so, can the CSI feedback quality be improved by using more bandwidth in the low energy scenario?
In the second part of this paper (Section V), we assume that both the TX and the RX harvest energy. In this case, the 
transmission power policy and the feedback policy are coupled, and need to be jointly optimized. Results from multivariate 
majorization theory are used to devise simple algorithms. We start by giving a brief preliminary description of majorization 
theory in Section II. Numerical results are presented in Section VI to validate the analysis. 
Finally, Section VII concludes the paper.

\textbf{Notation}:
Boldface letters are used to denote matrices and vectors. The transpose and conjugate transpose of matrix 
$\mathbf{A}$ is denoted by $\mathbf{A}^{\Tr}$ and $\mathbf{A}^{\He}$, respectively. We use $d_{i,j}$ to 
denote the element at the $i$-th row and $j$-th column of matrix $\mathbf{D}$, and $|\mathcal{S}|$ to denote 
the cardinality of the set $\mathcal{S}$. The set of integers from $m$ to $n$, $m<n$, is represented by $\left[m:n\right]$.
The algorithm with name ``Algo''  is represented as [output arguments]= Algo (input arguments).  
A circularly-symmetric complex Gaussian distributed random variable $\eta$ with zero mean and variance $\sigma^2$ is 
denoted by $\eta\sim \mathcal{CN}(0,\sigma^2)$.


 \section{PRELIMINARIES} \label{sec2}
In this section, the basic notion of majorization is introduced and some important inequalities on convex functions 
that are used in this work are stated.
The readers are referred to \cite{edmundson}, \cite{marshall} for a complete reference.
We start by stating the Edmundson-Madansky's inequality.
\begin{theorem}\cite{edmundson}
\label{th_one}
If $f$ is a convex function and $x$ is a random variable with values in an interval $[a,b]$, then
\begin{equation*}
\Exp \left[f\left(x\right)\right] \leq \frac{b-\mu}{b-a} f\left(a\right)+\frac{\mu-a}{b-a} f\left(b\right),
\end{equation*}
where $\mu$ is the mean of $x$.
\end{theorem}

Majorization theory formalizes the 
notion that the components of a vector $\bm{x}$ are ``less spread out'' than the components of a vector $\bm{y}$.

\begin{definition}
\label{def_2.1}
~Let $\bm{x}=\left[x_{1},\ldots, x_{n}\right], \bm{y}=\left[y_{1},\ldots, y_{n}\right]$, $\bm{x}, \bm{y}\in \mathbb R^{n}$ 
and let $x_{\left(i\right)}$ denote the $i$-th largest component of $\bm{x}$. Then $\bm{x}$ is said to be \textit{majorized} 
by $\bm{y}$, denoted by $\bm{x}\preceq\bm{y}$, if 
\begin{equation*}
\begin{aligned}
    \sum_{i=1}^l x_{\left(i\right)} &\le \sum_{i=1}^l y_{\left(i\right)}, \qquad \forall l \in \left[1:n-1\right]\\
		\sum_{i=1}^n x_{\left(i\right)} &= \sum_{i=1}^n y_{\left(i\right)}.
\end{aligned}
\end{equation*}
\end{definition}

\begin{definition}\cite[2.A.1]{marshall}
\label{def_2.2}
~ An $n \times n$ matrix $\mathbf{D}$ with elements $d_{i,j}$ is \textit{doubly stochastic} if
\begin{equation*}
\begin{aligned}
	&d_{i,j} \geq 0, \qquad \forall i, j \in \left[1:n\right], \\
   &\sum_{i=1}^n d_{i,j}=1, ~\forall j \in \left[1:n\right] ~\text{and}~ \sum_{j=1}^n d_{i,j}=1,~\forall i \in \left[1:n\right].
		\end{aligned}
		\end{equation*}
\end{definition}

\begin{theorem}\cite[4.A.1, 4.B.1]{marshall}
\label{thm_2.1}
For $\bm{x}, \bm{y} \in \mathbb R^{n}$, the following conditions are equivalent:
\begin{itemize}
\item $\bm{x}\preceq\bm{y}$.
\item $\bm{x} = \bm{y} \mathbf{D}$ for some doubly stochastic matrix $\mathbf{D}$.
\item For all continuous concave functions $g: \mathbb R \rightarrow \mathbb R$, $\sum_{i=1}^n g\left(x_i\right) 
\geq \sum_{i=1}^n g\left(y_i\right)$.
\end{itemize}
\end{theorem}

\begin{definition}\cite[15.A.2]{marshall}
\label{def_2.3}
~Let $\mathbf{X}$ and $\mathbf{Y}$ be $m \times n$ real matrices. Then $\mathbf{X}$ is said to be \textit{majorized} 
by $\mathbf{Y}$, written $\mathbf{X}\preceq \mathbf{Y}$, if $\mathbf{X}=\mathbf{Y}\mathbf{D}$, where the $n\times n$ 
matrix $\mathbf{D}$ is doubly stochastic.
\end{definition}

\begin{theorem}\cite[15.A.4]{marshall}
\label{thm_2.2}
~Let $\mathbf{X}$ and $\mathbf{Y}$ be $m \times n$ real matrices. Then, $~\mathbf{X} \preceq \mathbf{Y}$ if and only if 
$$
\sum\limits_{i=1}^n g\left(\bm{x}^{c}_i\right) \geq \sum\limits_{i=1}^n g\left(\bm{y}^{c}_i\right), 
$$ 
for all continuous concave functions $g: \mathbb R^{m} \rightarrow \mathbb R$; here $\bm{x}^{c}_i$ and $\bm{y}^{c}_i$ 
denote the $i$-th column vector of $\mathbf{X}$ and $\mathbf{Y}$, respectively.
\end{theorem}


 \section{System model} \label{sec3}
\begin{figure}
\centering
\vspace{-0.05\columnwidth}
\includegraphics[width=1.3\columnwidth,height=0.42\textheight] {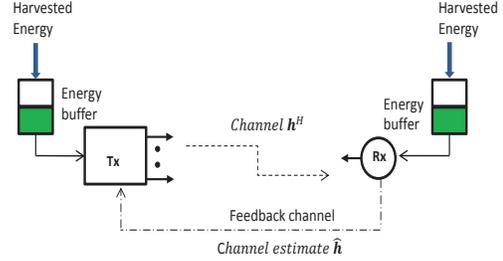}
\vspace{-0.2\columnwidth}
\vspace{-0.55\columnwidth}
\caption{MISO channel with feedback, where both the TX and the RX harvest and store ambient energy.}
 \label{fig_3.1}
\end{figure}

We consider a p2p MISO fading channel as shown in Fig.~\ref{fig_3.1}, where both the TX and the RX harvest energy from 
the environment. Each node is equipped with an individual energy buffer, i.e., a rechargeable battery, that can store 
the locally harvested energy.
\subsection{Energy Harvesting Model}
The total observation time is divided into $K$ equal length EH intervals. 
At the beginning of the $k$-th EH interval, $k \in [1: K]$, energy packets of size $e^{t}_{k}, e^{r}_{k}$ 
units arrive at the TX and the RX, respectively. At each node, this energy is first stored in an infinite 
size energy buffer, and used only for communication purposes, i.e., TX sending data, and the RX feeding back the CSI. 
We assume that all $e^{t}_{k}, e^{r}_{k}$'s are known in advance by both terminals. This model is suitable for an 
EH system in which the time-varying harvested energy can be accurately predicted \cite{deniz_tuto}. 

\subsection{Communication System Model}
\begin{figure}
\centering
\vspace{-0.1\columnwidth}
\includegraphics[width=1\columnwidth,height=0.4\textheight] {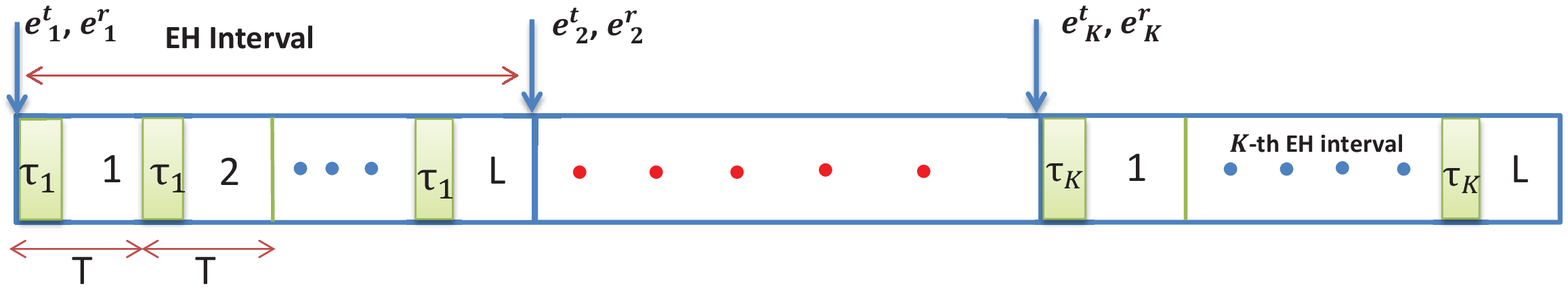}
\vspace{-0.8\columnwidth}
\caption{Energy harvesting time frame structure.}
 \label{fig_3.2}
\end{figure}

 Each EH interval consists of $L$ data frames, each of length $T$ channel uses. We assume a block fading 
channel model. The channel is constant during $T$ channel uses of each frame, but changes in an independent 
and identically distributed (i.i.d.) fashion from one frame to another. The time frame structure is shown in 
Fig.~\ref{fig_3.2}. The TX has $M>1$ antennas, while the RX has a single antenna. The received signal in a given channel 
use is given by
\begin{equation}
y= {\bm{h}^{\He}} \bm{w} s +\eta,
\label{eqn_3.1}
\end{equation}
where ${\bm{h}} \in \mathbb{C}^{M \times 1}$ represents the vector of channel coefficients from TX to the RX 
with i.i.d. $\mathcal{CN}(0,1)$ elements, $\bm{w}\in \mathbb{C}^{M \times 1}$ denotes the beamforming vector, 
the input symbol maximizing the achievable ergodic rate in the $k$-th EH interval is $s \sim \mathcal{CN}(0,p_k)$, 
and $\eta\sim \mathcal{CN}(0,1)$ represents the noise at the RX. 

\subsection{Feedback Model}
We assume that the RX perfectly estimates the channel state at the beginning of each data frame, and feeds 
back the quantized CSI to the TX within the same frame. In the $k$-th EH interval, the frame structure is 
as follows: The RX in $\tau_k$ channel uses sends the CSI through a feedback channel (uplink) which is 
modeled as an additive white Gaussian noise (AWGN) channel. In the remaining $T-\tau_k$ channel uses, TX sends 
data to the RX (downlink) exploiting the obtained CSI. The feedback model represents the Time-Division Duplex (TDD) 
system in which uplink and downlink use the same band in a time-sharing fashion, but the communication devices are 
not self-calibrated, and hence, induce non-reciprocal effects \cite{Santi_10}, \cite{Mari_11}.
In the above model, although the feedback overhead incurs a cost in the downlink bandwidth, a similar trade-off 
in the resource allocation between the CSI feedback quality and uplink data rate also arise in a Frequency-Division 
Duplex (FDD) system \cite{Mari_11}. Hence, the analytical results obtained in this paper are applicable in general 
settings, and for instance, can be used to address the trade-off between CSI quality and effective data rate in an FDD system.

In the $k$-th EH interval, quantization of the channel state is performed 
using a codebook $\mathpzc{C}_k$ known at both the TX and RX. The receiver uses Random Vector Quantization (RVQ). 
The codebook consists of $M$-dimensional unit vectors $\mathpzc{C}_k \triangleq\left\{\bm{f}_1,\ldots,\bm{f}_{2^{b_k}}\right\}$, 
where $b_k$ is the number of bits used for quantization. The RX chooses the beamforming vector according to 
$\bm{w}_k=\underset{\bm{f} \in \mathpzc{C}_k }{\argmax} ~{|\tilde{\bm{h}}^{\He}\bm{f}|}^2$, where 
$\tilde{\bm{h}}\triangleq\frac{\bm{{h}}}{||\bm{h}||}$. We assume that the length of the EH interval is very large 
compared to the channel coherence time (i.e., $L$ is very large). As a result, the achievable ergodic rate in the $k$-th 
EH interval is given by
\begin{equation}\label{eqn_3.2}
\begin{aligned}
R_k&=\left(1-\frac{\tau_k}{T}\right) \Exp_{||\bm{h}||^2,\nu_k}\left[\log_2\left(1+\frac{p_k}{
\left(1-\frac{\tau_k}{T}\right)}\left\|\bm{h}\right\|^2 
 \nu_k \right)\right],
\end{aligned}
\end{equation}
where $\nu_k={|\tilde{\bm{h}}^{\He}\bm{w}_k|}^2$. Note that $\nu_k$ and $||\bm{h}||^2$ are independent \cite{love_07}.
By using the AWGN feedback channel model, the number of feedback bits $b_k$ can be related to the energy used by the RX, $q_k$, 
and the number of channel uses $\tau_k$ as follows:
\begin{equation}
\label{eqn_3.3}
b_{k}=\tau_k \log_2\left(1+\frac{q_k}{{\tau_k\sigma^2}}\right),
\end{equation}
where $\sigma^2$ is the noise variance in the uplink.
For analytical tractability, we neglect the practical constraint that $b_{k}$ should be an integer.
Using the ergodic rate expression given in \cite[Equation (27)] {love_07} and (\ref{eqn_3.3}), the ergodic rate 
$R_k \triangleq R\left(p_k, q_k, \tau_k\right)$ is found to be
\begin{equation}
\label{eqn_3.4}
\begin{aligned}
R_{k}&=\left(1-\frac{\tau_k}{T}\right) \log_2e  \left(  e^{\rho_k} \sum_{l=0}^{M-1} E_{l+1}\left(\rho_k\right) \right. -\\
&  \left. \int\limits_{\nu_k=0}^{1} \left(1-\left(1-\nu_k\right)^{M-1}\right)^{N_k} \frac{M}{\nu_k} e^{\left(\frac{\rho_k}{\nu_k}\right)} 
E_{M+1}\left(\frac{\rho_k}{\nu_k}\right) d\nu_k \right)
\end{aligned}
\end{equation}
where  $\rho_k=\left(\frac{1-\frac{\tau_k}{T}}{p_k}\right)$, $N_k=\left(1+\frac{q_k}{\tau_k\sigma^2}\right)^{\tau_k}$, 
and $E_n\left(x\right)\triangleq \int_{1}^{\infty} e^{-xt}x^{-n} dt$ is the $n$-th order exponential integral.
\subsection{Optimization Problem}
The problem of maximizing the sum throughput by the end of the $K$-th EH interval can be formulated as

\begin{subequations}
\begin{align} \label{eqn_3.5}
&\underset{p_k, q_k, \tau_k}{\max} && \sum\limits_{k=1}^K R_{k} \\\label{eqn_3.5.1}
&\text{~s.t.} && L\sum\limits_{i=1}^l q_i \leq \sum\limits_{i=1}^l e^{r}_{i},~  \forall l \in [1:K], \\\label{eqn_3.5.2}
&&& LT\sum\limits_{i=1}^l p_i \leq \sum\limits_{i=1}^l e^{t}_{i},~ \forall l \in [1:K], \\\label{eqn_3.5.3}
&&&\tau_k\in[0,T),~p_k\geq0,~\text{and}~ q_k\geq0,\forall k \in [1:K].
\end{align}
\end{subequations}
The constraints (\ref{eqn_3.5.1}) and (\ref{eqn_3.5.2}) guarantee the \textit{energy neutrality} of the system, i.e., 
at each node, energy consumed can not be more than the energy harvested till that time.
Also note that $\tau_k$ impacts the achievable rate $R_k$ in each EH interval.

Coming up with simple algorithms to solve the optimization problem is desirable in 
EH networks as the nodes may not have the computational and energy resources for running complex optimization 
algorithms. However, the ergodic rate expression used in the above optimization problem is not in closed form 
and offers little insight into the convexity of the problem which is required to reduce the complexity of 
optimization. This motivates the use of convex bounds on (\ref{eqn_3.4}) as the objective function in the following 
optimization problems. Solving these modified problems provides an upper bound on the throughput. Since the constraints 
in the original and the modified optimization problems are the same, the solution for the modified problem is
also feasible in the original problem, and if used in evaluating the exact rate expression in (\ref{eqn_3.4}), 
we obtain a lower bound on the throughput. In some settings, we show that the bounds used are very close to the ergodic rate.

Before tackling the above problem, first, we consider a special case in which only the RX harvests energy. 
Later, the general case with both the TX and the RX harvesting energy is studied.


\section{EH Receiver}
In this setting, the RX harvests energy from the environment, whereas the TX is connected to the power grid 
so that it has a fixed power supply at all times. Therefore, there are no EH constraints at the TX, and constraints 
(\ref{eqn_3.5.2}) can be ignored. However, there is now a constraint on the average transmission power at each data 
frame of the $k$-th EH interval i.e., $p_k \leq p, \forall k$. 
The expected value $\nu_k$ is given by \cite{love_07,jindal}
\begin{equation}\label{exp_val}
\Exp [\nu_k]=1-2^{b_k}\beta\left(2^{b_k},\frac{M}{M-1}\right),
\end{equation}
where $\beta\left(x,y\right)$ denotes the beta function.
Using the quantization error bound in \cite[Lemma 6]{jindal}, (\ref{exp_val}) can be bounded 
as\footnote{This bound is universal in the sense that it applies to any $b_k$-bit quantization of an isotropically 
distributed vector, not necessarily limited to RVQ.}
\begin{equation}
\label{qt_bound}
\Exp [\nu_k] \leq \nu^{u}_{k}\triangleq{1-\left(\frac{M-1}{M}\right) 2^{\frac{-b_k}{M-1}}}.
\end{equation}
Applying Jensen's inequality on (\ref{eqn_3.2}), substituting (\ref{qt_bound}) and (\ref{eqn_3.3}), and using 
the fact that $\Exp {\|\bm{h}\|^2}=M$,  
an upper bound on the ergodic rate $R^{u}_{k} \triangleq R^{u}\left(p_k, q_k, \tau_k\right)$ is obtained as
\begin{equation}
\label{eqn_3.7}
R^{u}_{k}=t_k \log_2\left[1+\frac{p_k M}{t_k}{\left(1-\frac{M-1}{M}\left(1+\frac{q_k}{\tau_k\sigma^2}\right)^
{\frac{-\tau_k}{M-1}}\right)}\right],
\end{equation}
where $t_k \triangleq \left(1-\frac{\tau_k}{T}\right)$.

We now illustrate the tightness of the upper bound.
Applying the Jensen's inequality on (\ref{eqn_3.2}), $R^{u}_{k}-R_k$ can be lower bounded as
\begin{equation}\label{lb_1}
\begin{aligned}
R^{u}_{k}-R_k \geq t_k  &\log_2\left(1+\frac{p_k}{t_k}M 
 \nu^{u}_k \right) - \\
&t_k\Exp_{||\bm{h}||^2}\log_2\left(1+\frac{p_k}{t_k}\left\|\bm{h}\right\|^2 
 \Exp[\nu_k] \right).
\end{aligned}
\end{equation}
Since (\ref{eqn_3.2}) is a concave function of $\nu_k$ and $\nu_k \in \left[0,1\right]$, applying Theorem 
\ref{th_one} on (\ref{eqn_3.2}), we have
\begin{equation}\label{e_lb_1}
R_k \geq t_k\Exp_{||\bm{h}||^2}\log_2\left(1+\frac{p_k}{t_k}\left\|\bm{h}\right\|^2 
  \right) \Exp[\nu_k] 
\end{equation}
Now using (\ref{e_lb_1}), $R^{u}_{k}-R_k$ can be upper bounded as
\begin{equation}\label{lb_2}
\begin{aligned}
R^{u}_{k}-R_k \leq t_k&\log_2\left(1+\frac{p_k}{t_k}M \nu^{u}_k \right)- \\
&t_k\Exp_{||\bm{h}||^2}\log_2\left(1+\frac{p_k}{t_k}\left\|\bm{h}\right\|^2 
  \right) \Exp[\nu_k] 
\end{aligned}
\end{equation}
Since both $\lim_{b_k \to \infty}\nu^{u}_k=1$ and $\lim_{b_k \to \infty}\Exp[\nu_k]=1$ \cite{love_07},
and using (\ref{lb_1}) and (\ref{lb_2}), we have,
\begin{equation}\label{lb_3}
 \Delta R_k \triangleq \lim_{b_k \to \infty} R^{u}_{k}-R_k =t_k \Exp_{||\bm{h}||^2} \log_2\left(\frac{t_k+p_kM}
{t_k+p_k\left\|\bm{h}\right\|^2 }\right).
\end{equation}
Further, for all feasible $\tau_k$, in the low power regime,
\begin{equation}\label{ebb_1}
\lim_{p_k \to 0} \Delta R_k =0,
\end{equation}
and in the high power regime,
\begin{equation}\label{ebb_2}
\begin{aligned}
\lim_{p_k \to \infty} \Delta R_k&=t_k\left(\log_2M-\Exp_{||\bm{h}||^2}\log_2||\bm{h}||^2\right)\\
& \leq \log_2M-\Exp_{||\bm{h}||^2}\log_2||\bm{h}||^2.
\end{aligned}
\end{equation}
From the above analysis, it can be seen that when the RX has enough harvested energy to send large number of 
feedback bits, in the low power regime the bound is tight, and in the high power regime the difference is bounded 
by a constant. For example, it is $0.1958$ for $M=4$, and also note that $\lim_{M \to \infty} 
\log_2M-\Exp_{||\bm{h}||^2}\log_2||\bm{h}||^2=0$.

Using (\ref{eqn_3.7}) as the objective function, the modified optimization problem can be written as follows,
\begin{subequations}\label{eqn_3.8}
\begin{align}\label{prob 1:12}
&\underset{p_k, q_k, \tau_k}{\max} && \mathcal{U}=\sum\limits_{k=1}^K R^{u}_{k} \\\label{eqn_3.8.1}
&\text{~s.t.} && L\sum\limits_{i=1}^l q_i \leq \sum\limits_{i=1}^l e^{r}_{i},  \forall l \in [1:K], \\\label{eqn_3.8.2}
&&&  p_k \leq  p, ~\mbox{ and } p_k \geq 0, ~ \forall k \in [1:K], \\\label{eqn_3.8.3}
&&&\tau_k\in[0,T), ~\mbox{ and } ~q_k\geq0,~~\forall k \in [1:K],
\end{align}
\end{subequations}
where $p$ is the power constraint at the transmitter.
 
As the objective function is monotonic in $q_k$ and $p_k$, the constraint in (\ref{eqn_3.8.1}) must be satisfied 
with equality for $l=K$, and the first constraint in (\ref{eqn_3.8.2}) must be satisfied with equality, i.e., 
$p_k=p, \forall k$; otherwise, we can always increase $q_K, p_k$, and hence, the objective function, without 
violating any constraints. Now it remains to optimize over the variables $q_k$ and $\tau_k$.

The feasible set is represented as 
\begin{equation}
\mathfrak{F}=\left\{\bm{q}, \bm{\tau}| q_k, \tau_k \ \text{satisfy (\ref{eqn_3.8.1}), (\ref{eqn_3.8.3})}~ \forall k \right\},
\end{equation}
 where $\bm{q}=\left[q_1,\dots,q_K\right]$ and $\bm{\tau}=\left[\tau_1,\dots,\tau_K\right]$. 
To show that the above problem is a convex optimization problem, we make use of the following lemma.
\begin{lemma}\label{lem_3.1}
If the function $f\left(x,t\right): \mathbb R^{2}_{+} \rightarrow \mathbb R_{+}$ is concave, and $g\left(y,z\right): 
\mathbb R^{2}_{+} \rightarrow \mathbb R_{+}$ is concave and monotonically increasing in each argument, then the function 
$h\left(x,y,t\right)=\left(1-\frac{t}{T}\right)g\left(\frac{y}{1-\frac{t}{T}},\frac{f\left(x,t\right)}{1-\frac{t}{T}}\right)$ 
is concave $\forall \left(x,y\right)\in \mathbb R^{2}_{+}, t\in \left[0, T\right)$.
\label{l1}
\end{lemma}
\begin{IEEEproof}
The proof is similar to that of showing the perspective of a concave function is concave. See Appendix.
\end{IEEEproof}

\begin{proposition} \label{prop_3.1}
The objective function of the optimization problem (\ref{eqn_3.8}) is concave.
 \end{proposition}
\begin{IEEEproof}
See Appendix.
\end{IEEEproof}

\begin{figure}
\centering
\includegraphics[width=0.8\columnwidth] {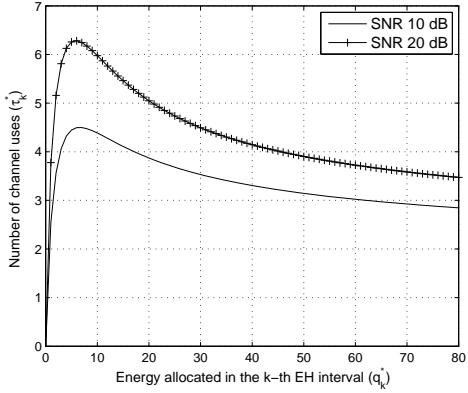}
\caption{Optimal number of channel uses for sending feedback.}
 \label{newfid}
\end{figure}

Since the objective function in (\ref{eqn_3.8}) is concave and the constraints are linear, it has a 
unique maximizer \cite{boyd}. Using the concavity of the objective function, we show that the optimal energy 
allocation vector is the most majorized feasible energy vector. 
\begin{proposition}\label{prop_3.2}
The global optimum of (\ref{eqn_3.8}) is obtained at $\left(\bm{q}^*,\bm{\tau}^*\right)$, 
where $\bm{q}^*\preceq\bm{q}, \forall \left(\bm{q},\bm{\tau}\right)\in \mathfrak{F}$, and $\tau^{*}_{k}$ 
is the solution of the following equation
\begin{equation}\label{eqn_3.9}
\frac{\partial R^{u}_{k}}{\partial \tau_k}|_{\left(q^{*}_{k},\tau^{*}_{k}\right)}=0, ~ \forall k\in \left[1:K\right].
\end{equation}
\end{proposition}
\begin{IEEEproof}
Consider the following equivalent form of (\ref{eqn_3.8}), where the optimization is performed in two steps.
\begin{equation}\label{mod_11}
\underset{ \bm{q}}{\max}~\mathcal{{\tilde{U}}}\left(\bm{q}\right) ~ \text{s.t.}~ \forall \left(\bm{q},\bm{\tau}\right) 
\in \mathfrak{F},
\end{equation}
where $\mathcal{{\tilde{U}}}\left(\bm{q}\right)$ is obtained by
\begin{equation}\label{eqn_3.10}
\mathcal{{\tilde{U}}}\left(\bm{q}\right)=\underset{\bm{\tau}}{\max}~{\mathcal{U}}\left(\bm{q},\bm{\tau}\right) 
~ \text{s.t.}~ \forall \left(\bm{q},\bm{\tau}\right) \in \mathfrak{F}.
\end{equation}

Since $\mathcal{U}$ is a concave function over the convex set $\mathfrak{F}$, the function 
$\mathcal{\tilde{U}}\left(\bm{q}\right)$ is concave, where the domain of $\mathcal{\tilde{U}}$ is the set 
$\mathfrak{\tilde{F}}=\left\{ \bm{q}| \left(\bm{q},\bm{\tau}\right) \in \mathfrak{F} \right\}$ \cite[3.2.5]{boyd}.  
$\mathcal{U}=\sum_{k=1}^K R^{u}_{k}$ is continuous, 
differentiable and concave in $\tau_k \in \left[0, T\right)$. Furthermore, for given $q_k$, $R^{u}_{k}$ 
approaches $\log_2\left(1+p\right)$ and $0$, as $\tau_k$ approaches $0$ and $T$, respectively. 
Therefore, the unique maximizer of (\ref{eqn_3.10}) lies in $\left[0, T\right)$, and it is obtained at
\begin{equation}\label{eqn_3.11}
\frac{\partial \mathcal{U}}{\partial \tau_k}|_{\tau^{*}_{k}}= \frac{\partial R^{u}_{k}}{\partial \tau_k}|_{\tau^{*}_{k}}=0, 
~ \forall k\in \left[1:K\right].
\end{equation}
From above, as $\tau^{*}_{k}$ is only a function of $q_k$, 
\begin{equation}\label{modif}
\mathcal{\tilde{U}}\left(\bm{q}\right)=\sum_{k=1}^K \tilde{R}^{u}_{k}
\end{equation}
where $\tilde{R}^{u}_{k} \triangleq \tilde{R}^{u}\left(q_k\right)={R}^{u}\left(q_k,\tau^{*}_{k}\left(q_k\right)\right)$.
Using (\ref{modif}) and Theorem \ref{thm_2.1}, $\mathcal{\tilde{U}}\left(\bm{q}^*\right)\geq \mathcal{\tilde{U}}\left(\bm{q}\right), 
\forall \bm{q} \in \tilde{\mathfrak{F}}$. Finding the optimal energy allocation vector $\bm{q}^*$ under the 
EH constraints turns out be a well known problem, and the algorithm to construct $\bm{q}^*$ is given in various 
works \cite{rui,zafer,ozel}. The proof that the algorithm constructs the most majorized feasible energy vector is 
given in \cite{ozel}. Since the optimal energy allocation vector is $\bm{q}^*$, the optimal $\bm{\tau}^*$ is obtained by (\ref{eqn_3.9}).
\end{IEEEproof}
A brief description of the algorithm tailored to this work is given next, while the details can be found in \cite{rui,zafer,ozel}. 
There is no closed form expression for the solution of (\ref{eqn_3.9}), hence we resort to numerical methods to obtain $\bm{\tau}^*$.
Fig.~\ref{newfid} shows the behavior of $\tau^{*}_{k}$ as a function of the allocated energy $q^{*}_{k}$. 

\subsection{Optimal Energy Allocation}\label{op_algo}
From Definition \ref{def_2.1}, we can see that the components of the most majorized energy vector are 
"less spread out" than any other feasible energy vector. Therefore, the algorithm essentially tries to make the 
energy vector as equalized as possible. This is done by spreading the energy to future intervals. However, note that the energy arriving in later intervals cannot be spread to earlier intervals due to the EH constraints.
The Optimal Energy Allocation (OEA) algorithm, given in Algorithm \ref{alg:tp}, divides the EH intervals 
into $\left|\mathcal{S}\right|$ energy bands whose indices form the set $\mathcal{S}=\left\{B_0, B_1,\ldots 
B_{\left|\mathcal{S}\right|}\right\}$, where $B_i <B_j, \forall i<j$, $B_0=0$, and 
$B_{\left|\mathcal{S}\right|}=K$. The $i$-th energy band contains the EH intervals with indices $k \in [B_{i-1}+1:B_i]$. 
Moreover, the optimal allocated energy values in each 
EH interval belonging to the $i$-th energy band are equal, and denoted by $q^{*}_{\left( i\right)}$.  
The energy vector $\bm{q}^*$ obtained by $\left[\bm{q}^*,\mathcal{S}_r\right]=\text{OEA}(K,\{e^{r}_{i}/L\})$, 
has the following properties:
\begin{itemize}
\item[{(P1)}] $q^{*}_k=q^{*}_{{\left(i\right)}}=\frac{\sum_{l=B_{i-1}+1}^{B_i} e^{r}_{l}}{L\left(B_{i}-B_{i-1}\right)}, 
~\forall k \in\left[B_{i-1}+1:B_{i}\right]$. 
\item[{(P2)}] The entries  $q^{*}_{{\left(i\right)}}$ are strictly monotonic, i.e., 
$q^{*}_{\left( 1\right)} < q^{*}_{\left( 2\right)} < ... < q^{*}_{\left( \left|\mathcal{S}\right|\right)}$.
\end{itemize}

\begin{algorithm}
\SetKwInOut{Input}{Input}\SetKwInOut{Output}{Output}
\Input{EH intervals $K$; Harvested energy $\{e_i\}$
}
\Output{Energy allocation $\bm{o}^{\star}$, Energy band indices $\mathcal{S}=\left\{B_0,B_1,\ldots 
B_{\left|\mathcal{S}\right|}\right\}$ } 
\BlankLine
\tcp{initialization}
$B_0 :=0$\;
\BlankLine
\For{$i=1:K$}
{

\For{$k=K:-1:(B_{i-1}+1)$}
    {
     (i) $o^{\star}_{l}=\frac{\sum_{j=B_{i-1}+1}^k e_j}{k-B_{i-1}}, \ l\in\left\{B_{i-1}+1,\ldots,k\right\}$ \\
    \If{$\sum_{i=1}^l o^{\star}_i \leq \sum_{i=1}^l e_i,  l=1,...,K$ }
		    { $B_i=k $;\\
			Save $\{o^{\star}_1, \cdots, o^{\star}_k\}$\\
		\textbf{break};}
       }
			    \If{$B_i==K$}
		 		{\textbf{break};}
		}
\caption{Optimal Energy Allocation (OEA) algorithm}\label{alg:tp}
\end{algorithm}


\section{EH Transmitter and Receiver}
In this section, we consider the general case where both the TX and the RX harvest energy. 
Note that if the TX is silent in the $k$-th interval, i.e., $p_k=0$, there is no incentive for the RX to send feedback in
this interval. Therefore, without loss of optimality we only consider EH profiles where $e^{t}_{1} > 0$. 
Otherwise, if there is an
EH profile such that $e^{t}_{k} = 0, k \in [1:m-1]$, then $p_k=0, k \in [1:m-1]$ due to the constraints in (\ref{eqn_3.5.2}).
In these intervals the RX simply accumulates the harvested energy, and without loss of optimality we can have a
new EH profile with $\tilde{e}^{t}_{1}=e^{t}_{i+m-1}, \forall i \in [1:K-m+1]$, and $\tilde{e}^{r}_{1}=\sum_{k=1}^{m}e^{r}_{k}$ and
$\tilde{e}^{r}_{i}=e^{r}_{i+m-1}, \forall i \in [2:K-m+1]$ for further analysis.

The ergodic rate upper bound in (\ref{eqn_3.7}) is not concave, 
but concave in each variable given the other variables are fixed. To obtain a simple algorithm and an upper 
bound on the throughput, we follow a similar approach as in the previous section, and use a 
concave upper bound on (\ref{eqn_3.7}) as the objective function for throughput optimization.

This bound is obtained by using a hypothetical system in which the transmission power is $1$ watt higher 
than the actual transmission power of the system, which is $p_k/t_k$. Plugging this into the upper bound in (\ref{eqn_3.7}),
a new upper bound $R^{ub}_{k} \triangleq R^{ub}\left(p_k, q_k, \tau_k\right)$ on the ergodic rate is obtained:
\begin{equation}
\label{eqn_3.12}
R^{ub}_{k}=t_k \log_2\left(1+\left(1+\frac{p_k}{t_k}\right)\frac{f_k}{t_k}\right),
\end{equation}
where $t_k \triangleq 1-\frac{\tau_k}{T}$ and $f_k\triangleq M-\left({M-1}\right)\left(1+\frac{q_k}{\tau_k\sigma^2}\right)
^{\frac{-\tau_k}{M-1}}$.
We now illustrate the tightness of the upper bound in (\ref{eqn_3.12}) in the low and high power regimes.
For all feasible $\tau_k, p_k$ and $q_k$, we can see that $0 < t_k \leq 1$ and $1 \leq f_k \leq M$. Consider
\begin{equation}\label{eqn_3.13}
\begin{aligned}
R^{ub}_k-R^{u}_k
&=t_k\log_2\left(\frac{t^{2}_k+t_k f_k+p_k f_k}{t_k+p_k f_k}\right)-t_k\log_2\left(t_k\right) \\
\end{aligned}
\end{equation}
Note that (\ref{eqn_3.13}) is decreasing in $p_k$ for fixed $\tau_k$ and $q_k$. Since $\tau_k, f_k$ are bounded, 
for fixed $\tau_k$ and $q_k$, in the low power regime 
\begin{equation}\label{pinf_2}
\begin{aligned}
\lim_{p_k \to 0} R^{ub}_k-R^{u}_k & =  t_k \log_2\left(1+\frac{f_k}{t_k}\right) \\
& \leq \log_2\left(1+M\right),
\end{aligned}
\end{equation}
and in the high power regime, 
\begin{equation}\label{eqn_3.14}
\begin{aligned}
\lim_{p_k \to \infty} R^{ub}_k-R^{u}_k& =- t_k \log_2({t_k}) \leq 0.5.
\end{aligned}
\end{equation}
From the above analysis, it can be seen that, (\ref{eqn_3.13}) decreases as the power is increased, 
and it is bounded by a constant in the high power regime.
By using (\ref{eqn_3.12}), the modified throughput maximization problem is formulated as
\begin{subequations}
\label{eq:SM_121}
\begin{align}\label{prob 1:121}
&\underset{p_k, q_k, \tau_k}{\max} && \mathcal{U}_1=\sum\limits_{k=1}^K R^{ub}_{k} \\\label{prob 1:111}
&\text{~s.t.} && L\sum\limits_{i=1}^l q_i \leq \sum\limits_{i=1}^l e^{r}_{i},  \forall l \in [1:K], \\\label{prob 1:411}
&&& LT\sum\limits_{i=1}^l p_i \leq \sum\limits_{i=1}^l e^{t}_{i}, \forall l \in [1:K], \\\label{prob 1:511}
&&&\tau_k\in[0,T),~p_k \geq 0,~q_k\geq0,~ \text{and} ~\forall k \in [1:K].
\end{align}
\end{subequations}
Since the objective function is monotonic in $q_k$ and $p_k$, the constraints in (\ref{prob 1:111}) and (\ref{prob 1:411}) 
must be satisfied with equality for $l=K$, otherwise, we can always increase $q_K, p_K$, and hence the objective function, 
without violating any constraints. The feasible set is represented as
\begin{equation*}
\mathfrak{J}=\left\{\left(\bm{p}, \bm{q}, \bm{\tau}\right)| p_k, q_k, \tau_k \ \text{satisfy (\ref{prob 1:111}), 
(\ref{prob 1:411}) and (\ref{prob 1:511})} ~\forall k\right\},
\end{equation*}
 where $\bm{p}=\left[p_1,\dots,p_K\right]$, $\bm{q}=\left[q_1,\dots,q_K\right]$ and $\bm{\tau}=\left[\tau_1,\dots,\tau_K\right]$. 
\begin{proposition}\label{prop_3.3}
The objective function in the optimization problem (\ref{eq:SM_121}) is concave.
\label{p2}
 \end{proposition}
\begin{IEEEproof}
See Appendix.
\end{IEEEproof}
Since the objective function in (\ref{eq:SM_121}) is concave and the constraints are linear, it has a 
unique maximizer \cite{boyd}. Consider the following equivalent form of (\ref{eq:SM_121}), where the 
optimization is performed in two steps.
\begin{equation}\label{mod_1}
\underset{\bm{p}, \bm{q}}{\max}~\mathcal{{\tilde{U}}}_1\left(\bm{p},\bm{q}\right) ~ \text{s.t.}~ \forall 
\left(\bm{p},\bm{q},\bm{\tau}\right) \in \mathfrak{J},
\end{equation}
where $\mathcal{{\tilde{U}}}_1\left(\bm{p},\bm{q}\right)$ is obtained by
\begin{equation}\label{mod_2}
\mathcal{{\tilde{U}}}_1\left(\bm{p},\bm{q}\right)=\underset{\bm{\tau}}{\max}~{\mathcal{U}}_1\left(\bm{p},\bm{q},\bm{\tau}\right)
 ~ \text{s.t.}~ \forall \left(\bm{p},\bm{q},\bm{\tau}\right) \in \mathfrak{J}.
\end{equation}
Since $\mathcal{U}_1$ is a concave function over the convex set $\mathfrak{J}$, the function $\mathcal{{\tilde{U}}}_1$ 
is concave with domain $\mathfrak{\tilde{J}}=\left\{ \left(\bm{p},\bm{q}\right)| \left(\bm{p},\bm{q},\bm{\tau}\right) 
\in \mathfrak{J} \right\}$ \cite[3.2.5]{boyd}. ${\mathcal{U}}_1=\sum_{k=1}^K R^{ub}_{k}$ is continuous, 
differentiable and concave in $\tau_k \in \left[0, T\right)$. Furthermore, for given $p_k$ and $q_k$, $R^{ub}_{k}$ 
approaches $\log_2\left(2+p_k\right)$ and $0$, as $\tau_k$ approaches $0$ and $T$, respectively. Therefore, 
the unique maximizer of (\ref{mod_2}), $\tau^{*}_{k},\forall k$ lies in $\left[0, T\right)$, and it is obtained as
\begin{equation}\label{a2}
\frac{\partial {\mathcal{U}}_1}{\partial \tau_k}|_{\tau^{*}_{k}}= \frac{\partial R^{ub}_{k}}
{\partial \tau_k}|_{\tau^{*}_{k}}=0,~ \forall k\in \left[1:K\right].
\end{equation}
As $\tau^{*}_{k}$ is only a function of $q_k$ and $p_k$, (\ref{mod_1}) can be written as
\begin{equation}
\begin{aligned}
&\underset{p_k, q_k}{\max}    
&&\mathcal{\tilde{U}}_1=\sum\limits_{k=1}^K \tilde{R}^{ub}_{k} ~~\text{s.t.}\  \forall k, \left(p_k, q_k\right)
\in \tilde{\mathfrak{J}},
\end{aligned}
\IEEEyessubnumber\label{a3}
\end{equation}
where $\tilde{R}^{ub}_{k} \triangleq \tilde{R}^{ub} \left(p_k,q_k\right)={R}^{ub}\left(p_k,q_k,\tau^{*}_{k}
\left(p_k,q_k\right)\right)$.

In order to get an insight on how the optimal solution of (\ref{mod_1}) may look like, consider a simple scenario 
in which there is only a sum power constraint at the TX and the RX, i.e., the constraints in (\ref{prob 1:111}), 
(\ref{prob 1:411}) has to be satisfied for only $l=K$. In this case, by Jensen's inequality, the uniform power 
allocation at the TX and the RX is optimal\footnote{ In this section, with slight abuse of terminology we use 
the terms RX power and RX energy interchangeably.}.
However, due to the EH constraints, this may not be feasible.
Using this intuition, we can see that the optimal policy tries to equalize the powers as much as possible, while
satisfying the EH constraints. Next, we consider the case in which the EH profiles at the TX and the RX are similar, 
and show that the optimization problem is considerably simplified.

\subsection{{Similar EH Profiles}}
The EH profiles are similar in the sense that the most majorized feasible vectors obtained from the EH profiles
of the TX and RX, $\bm{p}^*$ and $\bm{q}^*$, have the same structure, i.e., if $p^{*}_{i}=c_1, \forall i \in [m:n]$, then  
$q^{*}_{i}=c_2, \forall i \in [m:n]$ for some constants $c_1,c_2 \geq 0$. We now give a formal definition.

\begin{definition}\label{simil}
By using the OEA algorithm, let $\left[\bm{q}^*,\mathcal{S}_r\right]=\text{OEA}(K,\{e^{r}_{i}/L\})$ and 
$\left[\bm{p}^*,\mathcal{S}_t\right]=\text{OEA}(K,\{e^{t}_{i}/LT\})$. EH profiles at the TX and the RX are said 
to be \textit{similar} if $\mathcal{S}_r=\mathcal{S}_t$.
\end{definition}

From Section \ref{sec2}, we can see that the definition of majorization for the vector case does not directly 
extend to the matrix case. If OEA algorithm is used at the TX and RX separately, we get the most individually 
majorized power vectors, which in general may not be the optimal solution of (\ref{mod_1}). However, we now show 
that if the EH profiles are similar, the above mentioned approach is indeed optimal.

\begin{proposition}\label{prop_3.4}
If the EH profiles at the TX and the RX are similar then $\left(\bm{q}^*,\bm{p}^*,\bm{\tau}^*\right)$ is the global 
optimum of (\ref{eq:SM_121}), where $\bm{q}^*\preceq\bm{q}, \bm{p}^*\preceq\bm{p},~\forall \left(\bm{q},\bm{p},\bm{\tau}\right)
\in \mathfrak{J}$, and $\tau^{*}_{k}$ is the solution of 
\begin{equation}\label{partia}
\frac{\partial R^{ub}_{k}}{\partial \tau_k}|_{\left(p^{*}_{k},q^{*}_{k},\tau^{*}_{k}\right)}=0, ~ \forall k\in \left[1:K\right].
\end{equation}
\end{proposition}
\begin{IEEEproof}
See Appendix.
\end{IEEEproof}

\subsection{{Different EH Profiles}}
Unfortunately, we could not find a simple algorithm to solve (\ref{eq:SM_121}) in a general setting where 
the EH profiles are not similar. In (\ref{a3}), if one variable is fixed, optimizing over the other variable has 
a \textit{directional} or \textit{staircase water-filling} interpretation \cite{ulkus_11},\cite{rui}, however, 
the difficulty lies in the fact that there is no closed form expression for $\tilde{R}^{ub}_{k}$. 
Nonetheless, based on the convexity of the objective function, some properties of the optimal solution are given below.
\begin{lemma}\label{prop_3.5}
Under the optimal policy, the transmission power $p_k$, and the energy used to send the feedback $q_k$ are non-decreasing 
in $k, \forall k\in\left[1:K\right]$. 
\end{lemma}
\begin{lemma}\label{prop_3.6}
Under the optimal policy, at the time instants at which $R^{ub}$ changes, the energy buffer of either the 
TX or the RX is emptied.
\end{lemma}

The proofs of the above lemmas are given in Appendix.


\section{Numerical Results}
\begin{figure}
\centering
\includegraphics[width=0.8\columnwidth] {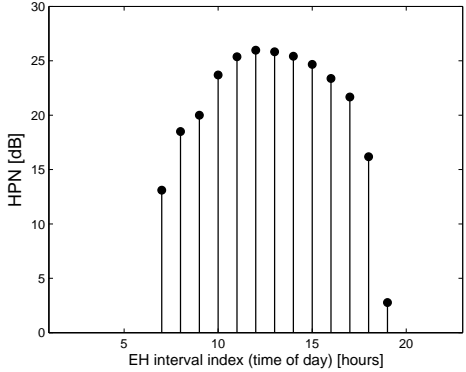}
\caption{Model for a solar energy harvesting profile.}
 \label{harv}
\end{figure}

We start by considering the case in which the RX harvests energy, while the TX has a constant power supply.
We assume that the RX is equipped with a solar EH device. Following \cite{reddy}, solar irradiance data 
is taken from the database reported in \cite{alexa:website}. Each EH interval is of duration $\Delta=1$ 
hour, $T=200$ ms, resulting in $L=18000$ frames. The harvested power from the irradiance data can be calculated 
as, $p_{harv}=I [\text{Watt}/m^2] \times Area [m^2] \times \rho$, where $\rho$ is the efficiency of the harvester.
A hypothetical solar panel of variable area is assumed. The area of the panel is adjusted such that we have the EH 
profile shown in Fig.~\ref{harv} at the RX. In Fig.~\ref{harv}, the harvested power to noise ratio (HPN) in each EH 
interval $\frac{e^{r}_k}{ \Delta \sigma^2}$ is shown. 

\begin{figure}
\centering
\centering
\includegraphics[width=0.9\columnwidth] {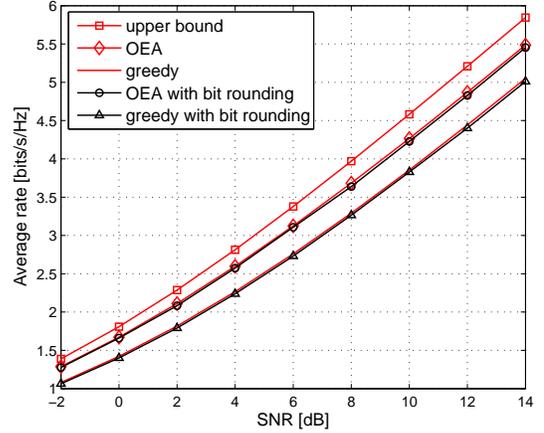}
\caption{Ergodic rate with only an EH RX, and $M=4$.}
 \label{erg_1}
\end{figure}

\begin{figure}
\centering
\includegraphics[width=0.8\columnwidth] {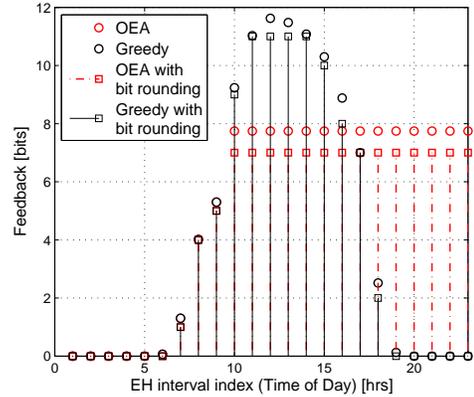}
\caption{Feedback load at downlink SNR of $10$ dB, $M=4$.}
 \label{bit}
\end{figure}

Using this EH profile, throughput of different feedback policies is shown in Fig.~\ref{erg_1}.
In Fig.~\ref{erg_1}, OEA represents the proposed policy in which the energy vector is obtained by
using the OEA algorithm, and then the optimal time span of feedback $\tau^{*}_{k}$ is obtained by solving (\ref{eqn_3.11}).
In the greedy scheme, the consumed energy is equal to the harvested energy in that interval, i.e., $q_k=e^{r}_k/L$,
and then optimization is performed only over ${\tau_k}$, given $q_k$. 
The performance of the above policies when the feedback bits are rounded
to the largest previous integer is also shown. We can see that the proposed approach outperforms the 
greedy policy by $1.6$ dB at a rate of $4$ bits/s/Hz. Also the rate loss due to 
bit rounding is negligible.
In Fig.~\ref{bit}, feedback bit allocation is shown for the above mentioned policies for a downlink SNR of $10$ dB. 
From Fig.~\ref{bit}, we can see that with the proposed strategy, feedback bit allocation is equalized as much as possible.

\begin{figure}
\centering
\centering
\includegraphics[width=0.85\columnwidth] {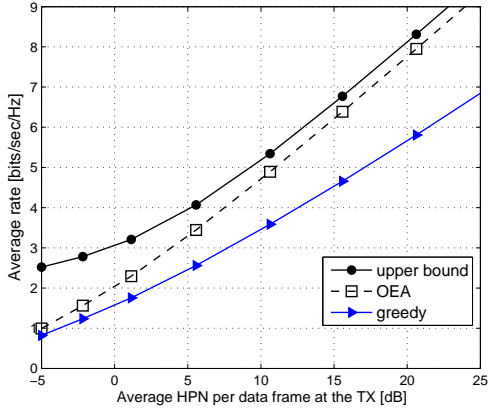}
\caption{Ergodic rate for similar EH profiles, $M=4$.}
 \label{erg_2}
\end{figure}

We now consider the case in which both the TX and the RX harvest energy, with similar EH profiles.
The same EH profile in Fig.~\ref{harv} is separately used at both the RX and the TX, hence the EH profiles are similar. 
In Fig.~\ref{erg_2}, the throughput of different schemes is shown at various mean HPN values at the TX.
The mean HPN at the TX is varied by increasing the harvester area at the TX, i.e., the EH profile is multiplied by 
a positive number (area), while keeping the same shape and efficiency.  
In Fig.~\ref{erg_2}, OEA represents the proposed policy in which the 
energy vector at the TX and the RX is obtained by using the OEA algorithm, and then the optimal time span of feedback $\tau^{*}_{k}$ 
is obtained by solving (\ref{a2}). 
In the greedy scheme, the allocated energy is equal to the harvested energy in that interval, i.e., at the
TX $p_k=e^{t}_k/LT$, at the RX $q_k=e^{r}_k/L$, and then optimization is performed only over ${\tau_k}$, 
given $p_k$ and $q_k$. 
The difference in throughput between the greedy and OEA is small when the average HPN is low, and it increases with the HPN.
In contrast to the OEA scheme, using the greedy approach with the solar EH profile results in some EH intervals 
being allocated zero energy, and therefore does not scale by increasing the harvester area.
This particularly hurts the greedy policy's throughput in the high HPN regime as the multiplexing gain (pre-log factor) is reduced.

\begin{figure}
\centering
\centering
\includegraphics[width=0.85\columnwidth] {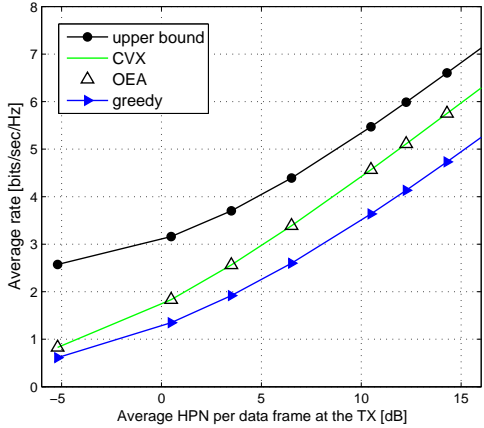}
\caption{Ergodic rate for non-similar EH profiles, $M=4$.}
 \label{erg_3}
\end{figure}

\begin{figure}
\centering
\includegraphics[width=0.9\columnwidth] {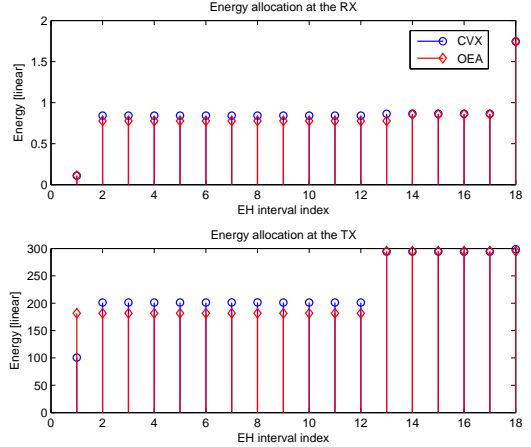}
\caption{Energy allocation at the TX and the RX, $M=4$.}
 \label{harv_4}
\end{figure}

Finally, we consider a case with non-similar EH profiles, where the EH profiles are generated independently at the 
TX and the RX, and they are i.i.d. with exponential distribution. 
EH profiles are verified not to be similar according to Definition \ref{simil}.
Similarly to Fig.~\ref{erg_2}, in Fig.~\ref{erg_3}, the mean HPN at the TX is varied by multiplying the 
EH profile by a constant, while keeping the same shape.
Since we could not find a simple algorithm in this case, CVX solver is used to solve the optimization problem \cite{boyd}, 
and is denoted as CVX in Fig.~\ref{erg_3}.
As we can see, the heuristic of using the OEA approach performs quite well even in the non-similar EH profile scenario.
The energy allocation at the TX and the RX are shown in Fig.~\ref{harv_4} for the above mentioned policies at an average 
per frame HPN of $0.5$ dB at the TX. Different from Fig.~\ref{erg_2}, in Fig.~\ref{erg_3} the rate scaling with 
average HPNs is same for both the greedy and the OEA policies. For the greedy policy, the allocated energy in an
EH interval scales with the increasing mean HPN, in contrast to the solar EH profile, for which the allocated
energy is zero in some intervals.


\section{Conclusion}
In contrast to the existing literature on the design of energy harvesting communication systems, we have assumed in this paper that the perfect channel state information is available only at the receiver side; and we have studied the problem of CSI feedback design in a p2p MISO channel under EH constraints at both the TX and the RX. Since the exact expressions of throughput are complicated, concave upper bounds 
have been used in the optimization problems.
We have first considered the case in which only the RX harvests energy, and optimized the feedback policy under EH constraints.
Later, the general case, in which both the TX and the RX harvest energy, is analyzed.
We have shown that, if EH profiles are similar, the optimization problem can be considerably simplified.
We remark that the result obtained in Proposition \ref{prop_3.4} is general, and, for example, it can be used 
in a network setting in which a concave utility is to be maximized in the presence of EH nodes with similar 
harvesting profiles and infinite size energy buffers. Numerical results show that the proposed policies 
not only outperform the greedy policy, but also achieve performances very close to the theoretical upper bound.  
Our work sheds light on the design of feedback-enabled multi-antenna systems when the nodes depend on EH devices for their energy.


\appendix

\subsection{Proof of Lemma \ref{lem_3.1}}
Let $X_1=\left[x_1\ y_1\ t_1\right]^{\Transpose}, X_2=\left[x_2\ y_2\ t_2\right]^{\Transpose}$, we have 
\begin{equation}
\begin{aligned}
&h\left(\lambda X_1+ \left(1-\lambda\right) X_2\right)\\
&=\Theta g\left(\frac{\lambda y_1+ \left(1-\lambda\right) y_2}{\Theta},\frac{ f\left(\overline{x}, \overline{t}\right)}{\Theta}\right)\\
&\stackrel{(a)}{\geq} \Theta g\left(\frac{\lambda y_1+ \left(1-\lambda\right) y_2}{\Theta},\frac{\lambda f_1+\left(1-\lambda\right) f_2}{\Theta}\right)\\
&=\Theta g\left(\frac{\Theta_1 y_1}{\Theta \alpha_1 }+ \frac{\Theta_2 y_2}{\Theta \alpha_2}, \frac{\Theta_1 f_1}{\Theta \alpha_1}+ \frac{\Theta_2 f_2}{\Theta \alpha_2} \right) \\
&\stackrel{(b)}{\geq}\Theta_1 g\left(\frac{y_1}{\alpha_1},\frac{f_1}{\alpha_1}\right)+
\Theta_2 g\left(\frac{y_2}{\alpha_2},\frac{f_2}{\alpha_2}\right)\\
&=\lambda h\left(X_1\right)+\left(1-\lambda\right) h\left(X_2\right),
\end{aligned}
\label{eq:SM_12}
\end{equation}
where $\overline{x} \triangleq \lambda x_1+\left(1-\lambda\right) x_2$, $\overline{t} \triangleq \lambda t_1+\left(1-\lambda\right) t_2$, $f_1\triangleq f\left( x_1, t_1\right)$, $f_2\triangleq f\left( x_2, t_2\right)$, $\Theta_1 \triangleq \lambda\left(1-\frac{t_1}{T}\right)$ and $\Theta_2 \triangleq \left(1-\lambda\right)\left(1-\frac{t_2}{T}\right), \Theta=\Theta_1+\Theta_2 $, $\alpha_1 \triangleq \left(1-\frac{t_1}{T}\right)$, $\alpha_2 \triangleq \left(1-\frac{t_2}{T}\right)$. Here
\begin{itemize}
\item[(a)] follows from the fact that $f\left(x,t\right)$ is concave, and $g\left(y,z\right)$ is monotonically increasing in each argument,
\item[(b)] follows from the fact that $\frac{\Theta_1}{\Theta}+\frac{\Theta_2}{\Theta}=1$, and $g\left(y,z\right)$ is concave.
\end{itemize}
\subsection{Proof of Proposition \ref{prop_3.1}}
Reproducing the ergodic rate bound in (\ref{eqn_3.7}) with $p_k=P,\forall k$, we have
\begin{equation}
R^{u}\left(q_k,\tau_k\right)=t_k\log_2\left(1+\frac{Pf_k}{t_k}\right),
\end{equation}
where $t_k\triangleq {1-\frac{\tau_k}{T}}, f_k\triangleq {(M-(M-1)(1+\frac{q_k}{\tau_k\sigma^2})^{\frac{-\tau_k}{M-1}})}$. 
Since $b_k$ in (\ref{eqn_3.3}) is concave in $q_k$ and $\tau_k$, it can be easily seen that $2^{-\frac{b_k}{M-1}}={\left(1+\frac{q_k}{\tau_k\sigma^2}\right)^{\frac{-\tau_k}{M-1}}}$ is convex, 
and hence, $f_k$ is concave. Using Lemma \ref{l1} with $g\left(y,z\right)=\log_2\left(1+z\right)$ and $f_k$,  
we can see that $R^{u}_{k}$ is concave. Since the objective function in (\ref{eqn_3.8}) is the summation of $R^{u}_{k}$'s, it is also concave.

\subsection{Proof of Proposition \ref{prop_3.3}}
First, we show that $g\left(y,z\right)=\log_2\left(1+\left(1+y\right) z\right), \left(y,z\right)\in \mathbb{R}^{2}_{+}$ is concave 
for $y\geq 0, z\geq 1$. The Hessian of $g$ is given by
\begin{equation}
\mathbf{J}=\frac{1}{\beta}
\begin{pmatrix}
	-z^2 & 1 \\
	1 & -\left(1+y\right)^2
 \end{pmatrix},
\end{equation}
where $\beta=\log_e2 \left(1+\left(1+y\right)z\right)^2> 0$. 
Consider $\bm{u}^{\Tr}\mathbf{J}\bm{u}=-\frac{1}{\beta}\left(a^2z^2+b^2\left(1+y\right)^2-2ab\right)$, where $\bm{u}=\left[a~b\right]^{\Tr} \in \mathbb R^{2}$. 
It can be easily seen that $\bm{u}^{\Tr}\mathbf{J}\bm{u} \leq 0$ for $ab\leq0$. For $ab>0$, since $z\left(1+y\right)\geq1$, $\bm{u}^{\Tr}\mathbf{J}\bm{u}=-\frac{1}{\beta}\left[\left(az-b\left(1+y\right) \right)^2+2ab\left(z\left(1+y\right)-1\right)\right] \leq 0$. As Hessian is negative semidefinite, $g\left(y,z\right)$ is concave.
Reproducing the ergodic rate bound in (\ref{eqn_3.12}), we have
\begin{equation}
R^{ub}_k= t_k \log_2\left(1+\left(1+\frac{p_k}{t_k}\right)\frac{f_k}{t_k}\right),
\end{equation}
where $t_k$ and $f_k$ are as defined before.

By following the similar steps in  Proposition \ref{prop_3.1}, $f_k$ can be shown to be concave.
Using Lemma \ref{l1} with $g\left(y,z\right)$ and $f_k$, 
we can see that $R^{ub}_{k}$ is concave. Since the objective function in (\ref{eq:SM_121}) is the summation of $R^{ub}_{k}$'s, it is also concave.

\subsection{Proof of Proposition \ref{prop_3.4} }
First, $\left({\bm{p}^*}, {\bm{q}^*}\right)$ is shown to be the solution of (\ref{a3}) and then $\bm{\tau}^*$ is obtained by (\ref{partia}).
Before solving (\ref{a3}), we prove that
\begin{equation}
\begin{aligned}
\left({\bm{p}^*}, {\bm{q}^*}\right)= &\underset{g,p_k, q_k}{\argmax} ~~ \sum\limits_{k=1}^K g\left(p_k,q_k\right) \\
& ~\text{s.t.}\  \forall k, \left(p_k,q_k\right)\in \tilde{\mathfrak{J}},g \in \mathfrak{C},
\end{aligned}
\IEEEyessubnumber\label{a4}
\end{equation}
where $\mathfrak{C}$ is the set of all continuous concave functions. As (\ref{a3}) is a special case of (\ref{a4}), $\left({\bm{p}^*}, {\bm{q}^*}\right)$ 
is also the solution of (\ref{a3}).

Before starting, we note that the notations and properties of the OEA algorithm discussed in Section \ref{op_algo} are used throughout the proof.
By contradiction, let us assume that there exists a ${[{\hat{\bm{p}}}^{\Tr} ~ {\hat{\bm{q}}}^{\Tr}]}^{\Tr} \neq  {[{\bm{p}^*}^{\Tr} ~ {\bm{q}^*}^{\Tr}]}^{\Tr}$ and $({\hat{\bm{p}}},{\hat{\bm{q}}})$ be the solution of (\ref{a4}). Then, by Theorem \ref{thm_2.2} we have,
\begin{equation}\label{prf_1}
{\left[{\hat{\bm{p}}}^{\Tr} ~ {\hat{\bm{q}}}^{\Tr}\right]}^{\Tr} \preceq  {\left[{\bm{p}}^{\Tr} ~ {\bm{q}}^{\Tr}\right]}^{\Tr}, ~ \forall \left(\bm{p},\bm{q}\right)\in \tilde{\mathfrak{J}}. 
\IEEEyessubnumber
\end{equation}
Since $\left(\bm{p}^*,\bm{q}^*\right)\in \tilde{\mathfrak{J}}$, by (\ref{prf_1}) and Definition \ref{def_2.3}, 
\begin{equation}\label{prf_2}
{\left[{\hat{\bm{p}}}^{\Tr} ~ {\hat{\bm{q}}}^{\Tr}\right]}^{\Tr} =  {\left[{\bm{p}^*}^{\Tr} ~ {\bm{q}^*}^{\Tr}\right]}^{\Tr} \mathbf{D}.
\end{equation} 
By the feasibility constraint in (\ref{prob 1:111}), 
\begin{equation}\label{prf_3}
\sum\limits_{j=B_{i-1}+1}^{B_{i}} {\hat{q}}_j \leq V_i = \sum\limits_{j=B_{i-1}+1}^{B_{i}} e^{r}_{j}/L,
\end{equation}
where $B_i$'s are the energy band indices as explained in Section \ref{op_algo}. 

Applying (\ref{prf_3}) for $i=1$, and remembering that $B_0=0$, we get 
\begin{equation}\label{prf_4}
\sum\limits_{j=1}^{B_{1}} {\hat{q}}_j  = \sum\limits_{j=1}^{B_{1}}\sum\limits_{i=1}^K  {q}^*_i d_{i,j} \leq V_1.
\end{equation}
By (P1) and (P2) in Section \ref{op_algo}, ${q}^*_i={q}^*_{\left(1\right)}+L_i$, where
\begin{equation}\label{prf_5}
\begin{aligned}
L_i&=0 ~~ \forall i\in\left[1:B_{1}\right], \\
L_i&>0 ~~ \forall i\in\left[B_{1}+1:K\right].
\end{aligned}
\end{equation}
From (\ref{prf_4}) and (\ref{prf_5})
\begin{equation}\label{prf_6}
\sum\limits_{j=1}^{B_{1}}\sum\limits_{i=1}^K  {q}^*_{\left(1\right)} d_{i,j} + \sum\limits_{j=1}^{B_{1}}\sum\limits_{i=B_{1}+1}^{K}  
{L}_i d_{i,j}  \leq V_1.
\end{equation}
Using the fact that $\mathbf{D}$ is doubly stochastic and by (P1), $B_1{q}^*_{\left(1\right)}=V_1$, and we have 
\begin{equation}\label{prf_7}
\sum\limits_{j=1}^{B_{1}}\sum\limits_{i=B_{1}+1}^{K}  {L}_i d_{i,j} \leq 0. 
\end{equation}
From (\ref{prf_5}) and (\ref{prf_7}), we get
\begin{equation}\label{prf_8}
d_{i,j}=0, ~~ \forall i\in\left[B_{1}+1:K\right], ~ \forall j\in\left[1:B_{1}\right].
\end{equation}
As $\mathbf{D}$ is doubly stochastic, using (P1) and (\ref{prf_8}), 
\begin{equation}\label{prf_9}
\hat{{q}}_j=\sum\limits_{i=1}^{B_{1}} {q}^*_{\left(1\right)} \sum\limits_{i=1}^{B_{1}} d_{i,j}={q}^*_{\left(1\right)} ={q}^*_{j} , 
\forall j\in\left[1:B_{1}\right].
\end{equation}
Since $\mathbf{D}$ is doubly stochastic, using (\ref{prf_8}), we get
\begin{equation}\label{prf_10}
\begin{aligned}
&\sum\limits_{i=1}^{B_{1}}\sum\limits_{j=1}^{K} d_{i,j} =B_1, 
&\sum\limits_{i=1}^{B_{1}}d_{i,j}=1, ~\forall j\in\left[1:B_{1}\right].
\end{aligned}
\end{equation}
We can rewrite (\ref{prf_10}) as
\begin{equation}\label{prf_11}
\begin{aligned}
\sum\limits_{i=1}^{B_{1}}\sum\limits_{j=1}^{K} d_{i,j}&=\sum\limits_{i=1}^{B_{1}}\sum\limits_{j=1}^{B_1} d_{i,j}+
\sum\limits_{i=1}^{B_{1}}\sum\limits_{j={B_1}+1}^{K} d_{i,j},\\
\end{aligned}
\end{equation}
from which it follows that
\begin{equation}\label{prf_t_1}
\sum\limits_{i=1}^{B_{1}}\sum\limits_{j={B_1}+1}^{K} d_{i,j}=0,
\end{equation}
and hence, 
\begin{equation}\label{prf_12}
\begin{aligned}
d_{i,j}=0, ~~ \forall i\in\left[1:B_{1}\right], \forall j\in\left[B_{1}+1:K\right].
\end{aligned}
\end{equation}

Then applying (\ref{prf_3}) for $i=2$,
\begin{equation}\label{prf_13}
\sum\limits_{j=B_{1}+1}^{B_{2}} {\hat{q}}_j = \sum\limits_{j=B_{1}+1}^{B_{2}} \sum\limits_{i=1}^K  {q}^*_i d_{i,j} \leq V_2.
\end{equation}

By (P1) and (P2), we have ${q}^*_i={q}^*_{\left(2\right)}+L_i$, where
\begin{equation}\label{prf_14}
\begin{aligned}
L_i&<0 ~~\forall i\in\left[1:B_{1}\right], \\
L_i&=0 ~~\forall i\in\left[B_{1}+1:B_{2}\right],\\
L_i&>0 ~~ \forall i\in\left[B_{2}+1:K\right].
\end{aligned}
\end{equation}
From (\ref{prf_13}) and (\ref{prf_14}),
\begin{equation}\label{prf_15}
\sum\limits_{j=B_{1}+1}^{B_{2}} \sum\limits_{i=1}^{K}  {L}_i d_{i,j} + \sum\limits_{j=B_{1}+1}^{B_{2}}
\sum\limits_{i=1}^K  {q}^*_{\left(2\right)} d_{i,j}  \leq V_2.
\end{equation}
Since $\mathbf{D}$ is doubly stochastic, by (P1), we obtain $\left(B_2-B_1\right){q}^*_{\left(2\right)}=V_2$, 
and using (\ref{prf_12}) and (\ref{prf_14}) in (\ref{prf_15}), we get
\begin{equation}\label{prf_16}
\sum\limits_{j=B_{1}+1}^{B_{2}} \sum\limits_{i=B_{2}+1}^{K}  {L}_i d_{i,j}   \leq 0 , L_i >0.
\end{equation}
From (\ref{prf_14}) and (\ref{prf_16}) it can be concluded that
\begin{equation}\label{prf_17}
d_{i,j}=0, ~~ \forall i\in\left[B_{2}+1:K\right],~ \forall j\in\left[B_{1}+1:B_{2}\right].
\end{equation}
As $\mathbf{D}$ is doubly stochastic, using (P1) together with (\ref{prf_12}) and (\ref{prf_17}), we have 
\begin{equation}\label{prf_18}
\hat{{q}}_j={q}^*_{\left(2\right)} \sum\limits_{i=B_{1}+1}^{B_{2}} d_{i,j} ={q}^*_{\left(2\right)}={q}^*_{j} , 
\forall j\in\left[B_{1}+1:B_{2}\right].
\end{equation}
Again, since $\mathbf{D}$ is doubly stochastic, using (\ref{prf_12}) and (\ref{prf_17}), 
\begin{equation}\label{prf_19}
\begin{aligned}
&\sum\limits_{i=B_{1}+1}^{B_{2}} \sum\limits_{j=1}^{K} d_{i,j} =B_2-B_1, \\
&\sum\limits_{i=B_{1}+1}^{B_{2}}d_{i,j}=1, ~\forall j\in\left[B_{1}+1:B_{2}\right].
\end{aligned}
\end{equation}
We can rewrite (\ref{prf_19}) as
\begin{equation}\label{prf_20}
\begin{aligned}
\sum\limits_{i=B_{1}+1}^{B_{2}}\sum\limits_{j=1}^{K} d_{i,j} &=\sum\limits_{i=B_{1}+1}^{B_{2}}\sum\limits_{j=B_{1}+1}^{B_{2}} d_{i,j} +
\sum\limits_{i=B_{1}+1}^{B_{2}}\sum\limits_{j={B_2}+1}^{K} d_{i,j}. \\
\end{aligned}
\end{equation}
From (\ref{prf_20}) we can see that
\begin{equation}\label{prf_n1} 
 \sum\limits_{i=B_{1}+1}^{B_{2}}\sum\limits_{j={B_2}+1}^{K} d_{i,j}=0,
\end{equation}
and hence,
\begin{equation}\label{prf_21}
d_{i,j}=0, \forall i\in\left[B_{1}+1:B_{2}\right]~\text{and} ~ \forall j\in\left[B_{2}+1:K\right].
\end{equation}

Continuing this approach for $i=3,..., \left( \left|\mathcal{S}_r\right|-1\right)$, we get $\hat{\bm{q}}=\bm{q}^*$. 
Since the EH profiles are similar, i.e., $\mathcal{S}_r=\mathcal{S}_t$, replacing $\hat{\bm{q}}$ by $\hat{\bm{p}}$ and $e^{r}_{j}$ by $e^{t}_{j}/T$ in 
the above proof, we reach the similar conclusion for $\hat{\bm{p}}$, i.e., $\hat{\bm{p}}=\bm{p}^*$. 
Therefore, ${[{\hat{\bm{p}}}^{\Tr} ~ {\hat{\bm{q}}}^{\Tr}]}^{\Tr} = {[{\bm{p}^*}^{\Tr} ~ {\bm{q}^*}^{\Tr}]}^{\Tr}$.

\subsection{Proof of Lemma \ref{prop_3.5} }
Assume that at least one of the $p_k, q_k$ is not monotonically increasing in $k$. Without loss of 
generality (w.l.o.s) we consider the cases in which $p_k>p_{k+1}, q_k \geq q_{k+1}$ and $p_k<p_{k+1}, q_k>q_{k+1}$. 
In the case of $p_k>p_{k+1}, q_k \geq q_{k+1}$, we can construct a new feasible policy, 
\begin{equation}\label{ord_1}
\begin{aligned}
\tilde{p}_{k}=\tilde{p}_{k+1}=\frac{p_k+p_{k+1}}{2}, \\
\tilde{q}_{k}=\tilde{q}_{k+1}=\frac{q_k+q_{k+1}}{2}.
\end{aligned}
\end{equation}
Since the objective function is concave, by Jensen's inequality, the new policy strictly increases the objective. 
Finally considering the case where $p_k<p_{k+1}, q_k>q_{k+1}$, we can construct another feasible policy, 
\begin{equation}
\begin{aligned}\label{od2}
\tilde{p}_{k}=p_k, ~\tilde{p}_{k+1}=p_{k+1},\\ 
\tilde{q}_{k}=q_{k+1}, ~\tilde{q}_{k+1}=q_{k}.
\end{aligned}
\end{equation}
The function $R^{ub}$ with variables $p, q,\tau$ can be written as,
\begin{equation}
R^{ub}\left(p, q,\tau\right) =t \log_2\left(1+\left(\frac{1}{t}+\frac{p}{t^2}\right) {f}\right),
\end{equation}
where $f \triangleq {M-\left(M-1\right)\left(1+\frac{q}{\tau\sigma^2}\right)^{\frac{-\tau}{M-1}}}$, $t \triangleq 1-\frac{\tau}{T}$ and $0\leq \tau < T$.
The second order partial derivative of $R^{ub}\left(p, q,\tau\right)$ is given by,
\begin{equation}\label{pdv}
\frac{\partial^2 R^{ub}}{\partial p \partial q}=\frac{\frac{\partial f}{\partial q}}{t\left(1+f/t+pf/t^2\right)^2}.
\end{equation}
Since $f$ is monotonic in $q$, (\ref{pdv}) is positive. As
 $\frac{\partial^2 R^{ub}}{\partial p \partial q} >0$, by the definition of derivative, 
\begin{equation}\label{pdv_1}
\begin{aligned}
R^{ub}\left(p, q,\tau\right)&+R^{ub}\left(p+\delta, q+\alpha,\tau\right) >\\
&R^{ub}\left(p+\delta, q,\tau\right)+R^{ub}\left(p, q+\alpha,\tau\right),~ \delta,\alpha >0.
\end{aligned}
\end{equation}

Since (\ref{pdv_1}) holds for all $0\leq \tau < T $, we have 
\begin{equation}\label{pdv_3}
\begin{aligned}
\tilde{R}^{ub} \left(p,q\right)+\tilde{R}^{ub} &\left(q+\delta,q+\alpha\right) > \\
&\tilde{R}^{ub} \left(p+\delta,q\right)+\tilde{R}^{ub} \left(p,q+\alpha\right),
\end{aligned}
\end{equation}
where $\tilde{R}^{ub}$ is obtained by,
\begin{equation}\label{pdv_4}
\tilde{R}^{ub} \left(p,q\right)=\underset{\tau}{\max} ~ R^{ub}\left(p, q,\tau\right).
\end{equation}
Finally, using (\ref{od2}) and (\ref{pdv_3}) we can see that the newly constructed policy strictly increases the objective.
\subsection{Proof of Lemma \ref{prop_3.6} }
Let us assume that the transmission rates in the $k$-th and the $k+1$-th intervals are different, i.e., $\tilde{R}^{ub}\left(p_k,q_k\right) \neq \tilde{R}^{ub}\left(p_{k+1},q_{k+1}\right)$. Before the $k+1$-th interval, the energy in the buffers of TX and the RX are $\Delta_r\triangleq \sum_{i=1}^k e^{r}_{i}- L\sum_{i=1}^k q_i$ and $\Delta_t\triangleq \sum_{i=1}^k e^{t}_{i}-LT\sum_{i=1}^k p_i$, respectively. W.l.o.s, we assume that $\Delta_r \leq \Delta_t$. We can construct another feasible policy 
\begin{equation}
\begin{aligned}\label{buf_1}
\tilde{p}_{k}=p_k+\delta, ~\tilde{p}_{k+1}=p_{k+1}-\delta,\\
\tilde{q}_{k}=q_{k}+\delta, ~\tilde{q}_{k+1}=q_{k+1}-\delta,
\end{aligned}
\end{equation}
where $\delta$ is chosen such that $\delta < \Delta_r$ and $\tilde{q}_{k} < \tilde{q}_{k+1}$. Now, (\ref{buf_1}) can be written as
 \begin{equation}
\begin{aligned}\label{buf_2}
\tilde{p}_{k}=\alpha p_k+\left(1-\alpha\right)p_{k+1},~\tilde{p}_{k+1}=\left(1-\alpha\right) p_k+\alpha p_{k+1},\\
\tilde{q}_{k}=\alpha q_k+\left(1-\alpha\right)q_{k+1},~\tilde{q}_{k+1}=\left(1-\alpha\right) q_k+\alpha q_{k+1},
\end{aligned}
\end{equation}
where $\alpha=1-\delta/\left(q_{k+1}-q_k\right)$.
Using Jensen's inequality
\begin{equation}\label{buf_3}
\sum_{j=k}^{k+1} \tilde{R}^{ub} \left(\tilde{p}_{j},\tilde{q}_{j}\right) >
\sum_{j=k}^{k+1}  \tilde{R}^{ub} \left(p_{j},q_{j}\right),
\end{equation}
which concludes the proof.

\bibliographystyle{IEEEtran}

\bibliography{EHMISO}

\begin{thebibliography}{10}
\providecommand{\url}[1]{#1}
\csname url@samestyle\endcsname
\providecommand{\newblock}{\relax}
\providecommand{\bibinfo}[2]{#2}
\providecommand{\BIBentrySTDinterwordspacing}{\spaceskip=0pt\relax}
\providecommand{\BIBentryALTinterwordstretchfactor}{4}
\providecommand{\BIBentryALTinterwordspacing}{\spaceskip=\fontdimen2\font plus
\BIBentryALTinterwordstretchfactor\fontdimen3\font minus
  \fontdimen4\font\relax}
\providecommand{\BIBforeignlanguage}[2]{{%
\expandafter\ifx\csname l@#1\endcsname\relax
\typeout{** WARNING: IEEEtran.bst: No hyphenation pattern has been}%
\typeout{** loaded for the language `#1'. Using the pattern for}%
\typeout{** the default language instead.}%
\else
\language=\csname l@#1\endcsname
\fi
#2}}
\providecommand{\BIBdecl}{\relax}
\BIBdecl

\bibitem{Kansal2007}
A.~Kansal, J.~Hsu, S.~Zahedi, and M.~B. Srivastava, ``{Power management in
  energy harvesting sensor networks},'' \emph{ACM Trans. Embed. Comput. Syst.},
  vol.~6, no.~4, Sept. 2007.

\bibitem{sudev}
S.~Sudevalayam and P.~Kulkarni, ``{Energy harvesting sensor nodes: survey and
  implications},'' \emph{IEEE Communications Surveys Tutorials}, vol.~13,
  no.~3, pp. 443--461, Mar. 2011.

\bibitem{ulkus_11}
O.~Ozel, K.~Tutuncuoglu, J.~Yang, S.~Ulukus, and A.~Yener, ``{Transmission with
  energy harvesting nodes in fading wireless channels: optimal policies},''
  \emph{IEEE JSAC}, vol.~29, no.~8, pp. 1732--1743, Sept. 2011.

\bibitem{elif}
M.~Antepli, E.~Uysal-Biyikoglu, and H.~Erkal, ``{Optimal packet scheduling on
  an energy harvesting broadcast link},'' \emph{IEEE JSAC}, vol.~29, no.~8, pp.
  1721--1731, Sept. 2011.

\bibitem{ulukus_bc}
J.~Yang, O.~Ozel, and S.~Ulukus, ``{Broadcasting with an energy harvesting
  rechargeable transmitter},'' \emph{IEEE Transactions on Wireless
  Communications}, vol.~11, no.~2, pp. 571--583, Feb. 2012.

\bibitem{deniz_leak}
B.~Devillers and D.~Gunduz, ``{A general framework for the optimization of
  energy harvesting communication systems with battery imperfections},''
  \emph{Journal of Comm. and Nets}, vol.~14, no.~2, pp. 130--139, Apr. 2012.

\bibitem{deniz_relay}
D.~Gunduz and B.~Devillers, ``{Two-hop communication with energy harvesting},''
  in \emph{CAMSAP}, San Juan, Puerto Rico, Dec. 2011.

\bibitem{haung}
C.~Huang, R.~Zhang, and S.~Cui, ``Throughput maximization for the gaussian
  relay channel with energy harvesting constraints,'' \emph{IEEE JSAC},
  vol.~31, no.~8, pp. 1469--1479, Aug. 2013.

\bibitem{deniz_tuto}
D.~Gunduz, K.~Stamatiou, N.~Michelusi, and M.~Zorzi, ``Designing intelligent
  energy harvesting communication systems,'' \emph{IEEE Communications
  Magazine}, vol.~52, no.~1, pp. 210--216, Jan. 2014.

\bibitem{love_08}
D.~Love, R.~Heath, V.~K.~N. Lau, D.~Gesbert, B.~Rao, and M.~Andrews, ``{An
  overview of limited feedback in wireless communication systems},'' \emph{IEEE
  JSAC}, vol.~26, no.~8, pp. 1341--1365, Oct. 2008.

\bibitem{edmundson}
H.~Edmundson, \emph{Bounds on the Expectation of a Convex Function of a Random
  Variable.}\hskip 1em plus 0.5em minus 0.4em\relax RAND Corporation, 1957.

\bibitem{marshall}
A.~W. Marshall, I.~Olkin, and B.~C. Arnold, \emph{Inequalities: Theory of
  majorization and its applications}.\hskip 1em plus 0.5em minus 0.4em\relax
  Springer, 2010.

\bibitem{Santi_10}
W.~Santipach and M.~Honig, ``Optimization of training and feedback overhead for
  beamforming over block fading channels,'' \emph{IEEE Trans on Information
  Theory,}, vol.~56, no.~12, pp. 6103--6115, Dec. 2010.

\bibitem{Mari_11}
M.~Kobayashi, N.~Jindal, and G.~Caire, ``{Training and feedback optimization
  for multiuser MIMO downlink},'' \emph{IEEE Transactions on Communications},
  vol.~59, no.~8, pp. 2228--2240, Aug. 2011.

\bibitem{love_07}
C.~K. Au-Yeung and D.~Love, ``{On the performance of random vector quantization
  limited feedback beamforming in a MISO system},'' \emph{IEEE Transactions on
  Wireless Commm}, vol.~6, no.~2, pp. 458--462, Feb. 2007.

\bibitem{jindal}
N.~Jindal, ``{MIMO} broadcast channels with finite-rate feedback,'' \emph{IEEE
  Trans on Inf. Theory}, vol.~52, no.~11, pp. 5045--5060, Nov. 2006.

\bibitem{boyd}
S.~Boyd and L.~Vandenberghe, \emph{{Convex optimization}}.\hskip 1em plus 0.5em
  minus 0.4em\relax New York, NY, USA: Cambridge University Press, 2004.

\bibitem{rui}
C.~K. Ho and R.~Zhang, ``{Optimal energy allocation for wireless communications
  with energy harvesting constraints},'' \emph{IEEE Transactions on Signal
  Processing}, vol.~60, no.~9, pp. 4808--4818, Sept. 2012.

\bibitem{zafer}
M.~A. Zafer and E.~Modiano, ``A calculus approach to energy-efficient data
  transmission with quality-of-service constraints,'' \emph{IEEE/ACM Trans.
  Netw.}, vol.~17, no.~3, pp. 898--911, Jun. 2009.

\bibitem{ozel}
O.~Ozel and S.~Ulukus, ``Achieving {{AWGN}} capacity under stochastic energy
  harvesting,'' \emph{IEEE Transactions on Information Theory}, vol.~58,
  no.~10, pp. 6471--6483, Oct. 2012.

\bibitem{reddy}
S.~Reddy and C.~Murthy, ``Dual-stage power management algorithms for energy
  harvesting sensors,'' \emph{IEEE Trans on Wireless Comm}, vol.~11, no.~4, pp.
  1434--1445, Apr. 2012.

\bibitem{alexa:website}
``{{Solar Resource and Meteorological Assessment Project (SOLRMAP)}},''
  \texttt{http://www.nrel.gov/midc/lmu/}.

\end{thebibliography}
\vspace*{-2\baselineskip}

\end{document}